\documentstyle[aps,prd,epsf]{revtex}

\newcommand{\be}{\begin{equation}}
\newcommand{\ee}{\end{equation}}
\newcommand{\bea}{\begin{eqnarray}}
\newcommand{\eea}{\end{eqnarray}}
\newcommand{\bdm}{\begin{displaymath}}
\newcommand{\edm}{\end{displaymath}}
\newcommand{\ul}{\underline}
\newcommand{\iinf}{\int_{S^2_{\infty}}}
\newcommand{\cod}{d^{\dagger}}
\newcommand{\we}{\wedge}
\newcommand{\sprod}[2]{\langle #1\, , \,#2 \rangle}
\newcommand{\EE}{\hat{E}}
\newcommand{\BB}{\hat{B}}
\newcommand{\Erpot}{{\cal E}}
\newcommand{\idid}{1 \! \! 1}

\newcommand{\RR}{\mbox{$I \! \! R$}}
\newcommand{\gtens}{\mbox{\boldmath $g$}}
\newcommand{\pmetric}{\mbox{\boldmath $p$}}
\newcommand{\Ptens}{{\cal P}}
\newcommand{\Qtens}{{\cal Q}}
\newcommand{\Mtens}{\Phi}
\newcommand{\dMtens}{d \Phi}
\newcommand{\dQtens}{d{\cal Q}}
\newcommand{\dPtens}{d{\cal P}}
\newcommand{\Rtens}{\mbox{\boldmath $R$}}
\newcommand{\Ttens}{\mbox{\boldmath $T$}}
\newcommand{\RPtens}{\Rtens ^{(p)}}

\begin{document}

\title{Mass Formulae for a Class of\\
Nonrotating Black Holes}

\author{Markus Heusler}

\address{Institute for Theoretical Physics \\
The University of Zurich \\
CH--8057 Zurich, Switzerland}

\date{\today}

\maketitle

\begin{abstract}
In the presence of a Killing symmetry, various 
self-gravitating field theories with massless scalars (moduli) 
and vector fields reduce to 
sigma-models, effectively coupled to $3$-dimensional gravity. 
We argue that this particular structure of the 
Einstein-matter equations gives rise to quadratic relations 
between the asymptotic flux integrals and the area and 
surface gravity (Hawking temperature) of the horizon. The 
method is first illustrated for the Einstein-Maxwell system. 
A derivation of the quadratic formula is then also presented 
for the Einstein-Maxwell-axion-dilaton model, which is 
relevant to the bosonic sector of heterotic string theory.
\end{abstract}

\section{Introduction}

It has been known for a long time that the Einstein-Hilbert
action in the presence of a Killing field, $k^{\mu}$, say,
describes a $2$-dimensional sigma-model effectively coupled
to $3$-dimensional gravity \cite{EH-59}. The target manifold 
of the sigma-model is the symmetric space $SL(2)/SO(2)$, 
which is parametrized in terms of two gravitational scalars 
(the norm of $k^{\mu}$ and its twist potential).

The Einstein-Maxwell (EM) system with a Killing symmetry
reveals a similar structure, where now the sigma-model
comprises the gravitational scalars and two additional 
electromagnetic potentials \cite{NK-69}. Again, the target 
manifold is a symmetric space, $G/H$. If the dimensional 
reduction is performed with respect to a timelike Killing 
field one finds $G/H = SU(2,1)/S(U(1,1) \times U(1))$, 
whereas $G/H = SU(2,1)/S(U(2) \times U(1))$ if $k^{\mu}$ is
spacelike. It is this particular property of the EM equations
which gives rise to the Ernst potentials \cite{E-68}, the 
Mazur identity \cite{PM-83} and -- in the presence of a 
second Killing field -- to the total integrability of the 
field equations \cite{DM-79}. Moreover, it is most likely 
that the black hole uniqueness theorem itself owes its 
existence to the sigma-model structure (see, e.g., \cite{MH-B}).

Obvious generalizations of the EM system are
self-gravitating field theories with massless scalars and 
Abelian vector fields. Considering scalar fields with 
symmetric target space $\bar{G}/\bar{H}$, Breitenlohner 
{\it et al.} \cite{BMG-88} were able to classify those models 
for which the dimensional reduction yields again a 
sigma-model with symmetric space $G/H$. Hence, these models 
admit a symmetry group which is large enough to comprise 
{\em all\/} scalar fields arising on the effective level within
{\em one\/} coset space. In terms of a representation $\Phi$ of
$G/H$, the field equations assume the form
\be
\RPtens \, = \, \mbox{Trace} \{ J \otimes J \} \, ,
\; \; \; \; 
d \ast J \, = \, 0 \, .
\label{i-1}
\ee
Here $\RPtens$ denotes the Ricci tensor with respect to the 
projection metric $\pmetric$, and $J$ is the sigma-model 
current,
\be
\pmetric \, \equiv \, V \gtens \, + \, k \otimes k \, ,
\; \; \; \; 
J \, \equiv \, \frac{1}{2} \, \Phi^{-1} d \Phi \, ,
\label{i-2}
\ee
where $\gtens$ is the spacetime metric and 
$V \equiv - g_{\mu \nu} k^{\mu} k^{\nu}$.

For the vacuum and the EM equations the explicit 
parametrization of the matrix $\Phi$ in terms of the target 
space coordinates (Ernst potentials) resulted form 
the work of Ehlers \cite{EH-59}, Ernst \cite{E-68}, Geroch 
\cite{BG-71}, Kinnersley {\it et al.} \cite{K-73-77}, 
\cite{KC-77}, Neugebauer and Kramer \cite{NK-69} and others.
Only recently, Gal'tsov and Kechkin were able to find the 
generalized Ernst potentials and the corresponding 
sigma-model representation for the 
Einstein-Maxwell-dilaton-axion (EMDA) equations \cite{GK-94}. 
The EMDA model is relevant to $N=4$ supergravity and to the 
bosonic sector of $4$-dimensional heterotic string theory. In
fact, this system provides the simplest (nontrivial) example 
of the models classified by Breitenlohner {\it et al.} 
\cite{BMG-88}. The relevant coset turns out to be
$Sp(4,\RR)/U(1,1)$, where the fact that $SO(2,3)$ is locally 
isomorphic to $Sp(4,\RR)$ is of crucial importance to the 
explicit representation of $\Phi$ \cite{DG-95}.

The matrix $J$ comprises $\mbox{dim}(G)$ algebraically 
independent current $1$-forms $j_{i}$, say. However, since the 
target manifold is a symmetric space, only $\mbox{dim}(G/H)$ 
of the conservation laws $d \ast j_{i} = 0$ are independent. 
By virtue of the Killing symmetry, each conserved current 
gives rise to a {\em closed\/} $2$-form 
$\Omega_{i} \equiv \ast (k \we j_{i})$. Integrating these 
$2$-forms over a spacelike hypersurface (which intersects the
horizon and extends to infinity), Stokes' theorem yields 
a set of relations between the asymptotic flux integrals, the 
corresponding horizon quantities and the values of the 
sigma-model fields (potentials) at the horizon.
In this way one obtains, for instance, the Smarr formula
\cite{LS-73} for stationary EM black hole configurations.

As one is only dealing with $\mbox{dim}(G/H)$ independent 
equations of the form $d \Omega_{i} = 0$, one might expect 
that Stokes' theorem yields as many relations between the 
charges and the horizon-values of the potentials. This is, 
however, not the case. In fact, the situation is better: 
Although there are $\mbox{dim}(H)$ conservation laws which 
can be obtained from the remaining ones, {\em all\/} 
currents $j_{i}$ are algebraically independent. For
this reason, Stokes' theorem yields $\mbox{dim}(G)$ 
nonredundant relations of the Smarr type when applied to the 
$2$-forms $\Omega_{i}$. The entire set of
relations may then be used to eliminate the unknown 
horizon-values of the sigma-model scalars. In this way one 
ends up with a relation which involves only the total charges 
and the corresponding horizon quantities. For both the EM and
the EMDA system we shall prove that all stationary black hole
configurations with nonrotating Killing horizon fulfill 
\be
M_{H}^{2} \, = \, M^{2} \, + \, N^{2} \, + \, 
D^{2} \, + \, A^{2} \, - \, Q^{2} \, - \, P^{2} \, ,
\label{i-3}
\ee
where the r.h.s. comprises the asymptotic flux integrals, 
i.e., the total mass, the NUT charge, the dilaton and axion 
charges, and the electric and magnetic charges, respectively. 
The quantity $M_{H}$ is the Komar integral over the horizon, 
$M_{H} = -(8 \pi)^{-1} \int_{H} \ast dk$. The 
left hand side (l.h.s.) of the 
above relation can therefore be expressed in terms of the 
area of the horizon, ${\cal A}$, and its surface gravity 
$\kappa$, or, equivalently, its Hawking temperature $T_{H}$,
\be
M_{H} \, = \, \frac{1}{4 \pi} \, \kappa \, {\cal A} \, = \, 
\frac{1}{2} \, T_{H} \, {\cal A} \, .
\label{i-4}
\ee

The ``extreme'' Reissner-Nordstr\"om solution is well known 
to fulfill the Bogomol'nyi bound $0 = M^{2} -Q^{2} - P^{2}$. 
The corresponding BPS bound for the EMDA system,
$0 = M^{2} + D^{2} + A^{2} - Q^{2} - P^{2}$, was obtained by
Cl\'ement and Gal'tsov \cite{CG-96}, constructing the null 
geodesics of the target space. Discussing the asymptotic 
behavior of target space geodesics for spherically symmetric 
configurations, Breitenlohner {\it et al.} obtained eq. 
(\ref{i-3}) with unspecified l.h.s. \cite{BMG-88}. 
In fact, many of the {\em spherically symmetric\/} black hole 
solutions with scalar and vector fields
(see, e.g., \cite{GG-82}, \cite{GM-88} and \cite{GHS-91}) 
are known to fulfill eq. (\ref{i-3}), where the l.h.s. is expressed in 
terms of the horizon radius (see also \cite{GL-96} and references
therein). Using the generalized first law of black hole
thermodynamics, Gibbons {\it et al.} \cite{GKK-96} were 
recently able to derive eq. (\ref{i-3}) for spherically 
symmetric solutions with an arbitrary number of vector and
moduli fields.

In the present paper we establish eq. (\ref{i-3}) for 
arbitrary soliton ($M_{H} \equiv 0$) and 
stationary, nonrotating black hole solutions of the EM and 
EMDA equations. Our derivation is not restricted to 
spherical symmetry, neither do we require 
the configurations to be static. The crucial observation is that
the coset structure gives rise to a set of Smarr type 
formulae which is sufficiently large to derive the desired 
relation. Since the EMDA sigma-model does not reduce to the 
EM sigma-model for vanishing dilaton and axion fields
\cite{CG-96}, we derive eq. (\ref{i-3}) separately for the 
two cases.

Although the recipe is simple, it is a rather unpleasant task
to write out the current matrix $J$ for a given representation
$\Phi$. We think 
that it should be possible to obtain the formula (\ref{i-3}) 
even without having an explicit representation of the
matrix $\Phi$ at hand. We therefore conjecture that
relations similar to eq. (\ref{i-3}) hold for all models which
reduce to the form (\ref{i-1}) in the 
presence of a stationary Killing symmetry.

The paper is organized as follows: We start with a simple 
example; the static, purely electric EM system. In this case, 
the conserved currents are derived ``from scratch'',  that is, 
without making use of the sigma-model structure (see also 
\cite{MH-B} for this approach). The third section 
is devoted to the general stationary EM equations. We recall 
the dimensional reduction and use the coset structure to
construct all conserved currents and 
closed $2$-forms. Integrating the latter 
over a spacelike hypersurface will provide us with a set of 
generalized Smarr formulae, which we then use to 
compute the horizon potentials and to
derive eq. (\ref{i-3}). In the fourth section the procedure 
is repeated for the EMDA system, where we take advantage of
the coset representation found by Gal'tsov and Kechkin \cite{GK-94}.
Since we prefer to use the exterior calculus, some 
computational rules for differential froms are compiled 
in the Appendix.

\section{A Simple Example}
As a motivation we consider the static, purely electric 
Einstein-Maxwell (EM) equations. 
In this case, the field strength $2$-form, $F = dA$, 
can be expressed in terms of the stationary Killing field 
($1$-form) $k$ and the electric $1$-form $E$: 
$F = \frac{k}{V} \we E$, where 
$V \equiv - k_{\mu} k^{\mu} \equiv - \sprod{k}{k}$.
Staticity implies that the twist of the Killing field vanishes
and therefore $d (k / V) = 0$ (see eq. (\ref{A-5})).
Hence, the Bianchi identity, $dF = 0$, and the Maxwell equation, 
$d \ast F = 0$, become
\be
dE \, = \, 0 \, , \; \; \; \; \;
\cod \left( \frac{E}{V} \right) \, = \, 0 \, ,
\label{Maxwell-1}
\ee
respectively, where $\cod = \ast d \ast$ 
denotes the coderivative operator. 
(Here we have used eq. (\ref{A-4}) for $\alpha = E/V$.)
In addition, we consider the ($00$)-component of Einstein's equations,
$\Rtens(k,k) = 8 \pi \Ttens(k,k) = \sprod{E}{E}$. In the static case, 
eq. (\ref{A-12}) reduces to the Poisson equation,
$\cod(dV/V) = -2\Rtens(k,k)/V$. 
Introducing the potential $\phi$, $d \phi = E$, and using 
the formula $\cod(f \alpha) = f \cod \alpha - \sprod{df}{\alpha}$ 
(for arbitrary functions $f$ and $1$-forms $\alpha$), 
eq. (\ref{Maxwell-1}) implies 
$\frac{1}{V} \sprod{E}{E} = -\cod \left({\phi}\frac{E}{V} \right)$.
Hence, both the Maxwell and the Poisson equation assume the form of 
current conservation laws:
\be
\cod j_Q = 0 \, , \; \; \; 
j_Q \, \equiv \, \frac{d \phi}{V} \, ,
\label{jQ-1}
\ee
\be
\cod j_M = 0 \, , \; \; \; 
j_M \, \equiv \, -\frac{1}{2} \frac{d V}{V} 
\, + \, \phi \frac{d \phi}{V} \, .
\label{jM-1}
\ee

In the presence of the Killing field $k$, every conserved $1$-form,
$j$, gives rise to a closed $2$-form, $\Omega \equiv \ast(k \we j)$.
As $d \Omega$ vanishes, Stokes' theorem, 
$\int_{\partial \Sigma}(\mbox{2-form}) =
\int_{\Sigma}d\, (\mbox{2-form})$, implies 
\be
\iinf \, \ast (k \we j) \, = \,
\int_H \, \ast (k \we j) \, ,
\label{Stokes-1}
\ee
where the integral on the r.h.s. extends over the topological 
$2$-sphere $H = {\cal H} \cap \Sigma$, ${\cal H}$ and $\Sigma$ 
being the horizon and a spacelike hypersurface, respectively. 
In order to apply this formula
to the above currents, we have to express $\ast(k \we j_Q)$
and $\ast(k \we j_M)$ in terms of the $2$-forms
$F$, $\ast F$ and $\ast dk$. This is
immediately achieved by using the static, purely electric
identities  $- (k \we dV/V) = dk$ and $(k \we E/V) = F$
(see eq. (\ref{A-5}). The closed
$2$-forms corresponding to the currents defined in
eqs. (\ref{jQ-1}) and (\ref{jM-1}) are
\be
\ast(k \we j_Q) \, = \, \ast F \, \; \; \; \mbox{and} \; \; 
\ast(k \we j_M) \, = \, \frac{1}{2}\ast dk \, + \, \phi \ast F \, ,
\label{JQJM}
\ee
respectively.

Defining the horizon quantities 
$M_H \equiv -\frac{1}{8 \pi} \int_H \ast dk$ and 
$Q_H \equiv -\frac{1}{4 \pi} \int_H \ast F$, 
and using the Komar expression
$M = -\frac{1}{8 \pi} \int_{\infty} \ast dk$
for the total mass of a stationary spacetime, 
as well as the corresponding
expression for the total charge,
$Q = -\frac{1}{4 \pi} \int_{\infty} \ast F$, we immediately 
find from eqs. (\ref{Stokes-1}) and (\ref{JQJM})
\be
Q \, = \ Q_H \, , \; \; \; \; \;
M \, = \, M_H \, + \, \phi_H \, Q_H \, ,
\label{relation-1}
\ee
which implies the Smarr formula, $M = M_H + \phi_{H} Q$. We also recall 
that -- for a Killing horizon ${\cal H}$ with null generator Killing 
field $k$ -- we have $M_H = \frac{1}{4 \pi} \kappa {\cal A}$, where 
$\kappa$ and ${\cal A}$ are, respectively, the surface gravity and 
the area of the horizon (at time $\Sigma$). Here we have adopted the 
gauge $\phi_{\infty} = 0$ and used the fact that the electric 
potential assumes a constant value on the Killing horizon, 
$\phi_H$, say. We also recall that asymptotic flatness and the 
Killing property of the horizon imply $V_{\infty} = 1$ and $V_H = 0$, 
respectively. As a consequence of the above relations (i.e., the
Smarr formula) the horizon-value of the 
electric potential is determined by the total mass $M$, 
the total charge $Q$ 
and the horizon quantities $\kappa$ and ${\cal A}$,
\be
\phi_H \, = \, \frac{1}{Q} \,  ( \, M \, - \, \frac{1}{4 \pi} 
\kappa {\cal A} \, ) \, .
\label{res-phi}
\ee

Until now we have only used Stokes' theorem and the fact 
that the field equations assume the form of differential 
conservation laws. One may wonder if 
there exist additional conserved currents which can be expressed in
terms of the $1$-forms $dV/V$ and $d \phi / V$. Although the
conservation laws for these currents will give rise to 
redundant equations on the differential level, they
may, nevertheless, provide us with new information after integration. 
This is due to the fact that the coefficients in front of the
$1$-forms $dV/V$ and $d \phi / V$ can be pulled out of the boundary 
integrals -- provided
that they depend only on the potentials $V$ and $\phi$, and assume 
therefore constant values on $H$ and $S^2_{\infty}$. 
In this way one obtains combinations of $M$, $Q$, $M_H$ and 
$Q_H$ which are independent 
on the relations (\ref{relation-1}) derived from the field 
equations. In fact, it is immediately verified from
eqs. (\ref{jQ-1}) and (\ref{jM-1}) that
\be
\cod j_3 = 0 \, , \; \; \; 
j_3 \, \equiv \,
\left(V + \phi^2 \right) \, \frac{d \phi}{V} \, - \, 
\phi \, \frac{d V}{V} \, .
\label{j1-1}
\ee
(Use $\cod(f \alpha) = f \cod \alpha - \sprod{df}{\alpha}$
(for arbitrary functions $f$ and $1$-forms $\alpha$) to show
that $j_3$ is conserved.)
We can therefore apply Stokes' theorem (\ref{Stokes-1}) to
the new closed $2$-form obtained from $j_3$,
\be
\ast (k \we j_3) \, = \, 
(V + \phi^2)  \ast F \, + \, \phi \ast dk \, .
\label{j1-2}
\ee
As the potentials assume constant values on the horizon and
at infinity, we immediately find 
$Q = \phi_{H}^2 Q_H + 2 \phi_{H} M_H = - \phi_{H}^2 Q + 2 \phi_{H} M$,
where we have also used eqs. (\ref{relation-1}) in the second step.
Now using the expression 
(\ref{res-phi}) for $\phi_H$ gives $Q^2 = (M - M_H)(M + M_H)$ and hence 
\be
M^2 \, = \, \left( \frac{1}{4 \pi} \kappa {\cal A} \right)^2 
\, + \, Q^2 \, \; \; \; \; \mbox{i.e.,} \; \; \; 
T_H\, = \, \frac{2}{{\cal A}} \sqrt{M^2 - Q^2} \, ,
\label{relation-2}
\ee
where $T_H = \frac{1}{2 \pi} \kappa$ is the Hawking temperature.

The relation between the charges and the horizon quantities 
following from eq. (\ref{j1-1}) was already derived by Israel 
in 1969 for a nondegenerate Killing horizon, $\kappa \neq 0$ 
\cite{WI-vac}. (The 
above derivation does not require that the horizon contains its 
bifurcation surface, implying that eq. (\ref{relation-2}) also 
holds in the degenerate case.) In fact, Israel and other authors
used quadratic relations of the above kind
to conclude that the electric potential depends only on the 
gravitational potential, $\phi = \phi(V)$. This important 
result opened the way for 
the extension of the vacuum Israel theorem \cite{WI-vac} to static 
electrovac black hole spacetimes \cite{WI-evac}, \cite{MZH}.

The existence of the additional conserved current (\ref{j1-1}) 
is not accidental: In the presence of a Killing field, the EM 
equations form a non-linear sigma-model (effectively coupled 
to $3$-dimensional 
gravity) with a symmetric target space $G/H$ \cite{NK-69} (see the 
next section). The isometries of the target space imply 
that -- besides the $\mbox{dim}(G/H)$ field equations -- there 
exists an additional set of 
$\mbox{dim}(H)$ conserved currents. In the static, purely 
electric case 
under consideration one ends up with the two equations 
(\ref{jQ-1}) and (\ref{jM-1}) for $V$ and $\phi$, respectively, 
and the additional conserved current $j_3$, given in 
eq. (\ref{j1-1}). 
(The {\em full\/} EM system comprises four plus four 
conserved currents; the truncation $U = 0$, $\psi = 0$ is, 
in this case, compatible with the coset representation. Here $U$ 
and $\psi$ are the twist and the magnetic potential, 
respectively, to be defined in the following section.)
 
\section{The Stationary Einstein-Maxwell System}

In the previous section we have restricted ourselves to the static, 
purely electric case. We shall now construct the complete set
of conserved currents for the general stationary EM equations.
We do so by taking advantage of the sigma-model structure of the
EM equations in the presence of a Killing field. The eight conserved
currents give rise to eight closed $2$-forms which will be
integrated over a spacelike hypersurface. The resulting Smarr
formulae are finally used to obtain the desired quadratic relation 
(\ref{i-3}) between the flux integrals and the quantity 
$M_H = \frac{1}{4\pi} \kappa {\cal A}$.

\subsection{Dimensional Reduction}

We start by briefly recalling some basic facts concerning the 
dimensional reduction of the Maxwell and the Einstein equations in 
the presence of a (stationary) Killing field \cite{BG-71} 
(see also \cite{IW-72}, \cite{KSMH-80} or \cite{MH-B}). Throughout 
this paper we use the symbols $k$, $V$ and $\omega$ for the 
stationary 
Killing field ($1$-form), its norm and its twist $1$-form, 
respectively:
\be
V \, \equiv \, - \sprod{k}{k} \, , \; \; \; \; \;
\omega \, \equiv \, \frac{1}{2} \ast \left( k \we dk \right) \, .
\label{defs}
\ee

In the presence of a Killing field, the Bianchi identity,
$dF = 0$, and the Maxwell equation, $d \ast F = 0$, 
give rise to two (local) scalar potentials $\phi$ and $\psi$, 
respectively: Since the Lie derivatives of $F$ and $\ast F$ 
with respect to $k$ vanish, one obtains 
(with $L_k = i_k \circ d + d \circ i_k$) the 
equations $d (i_k F) = 0$ and $d (i_k \ast F) = 0$, and hence
\be
E \, \equiv \, -i_k F \, = \, d \phi \, ,
\; \; \; \; \; 
B \, \equiv \, i_k \ast F \, = \, d \psi \, .
\label{Maxwell-2}
\ee
(Here and in the following $i_{k}\alpha$ denotes the interior 
derivative
of the $p$-form $\alpha$ with respect to $k$,
$(i_{k} \alpha)_{\mu_{2} \ldots \mu_{p}}
\equiv k^{\mu} \alpha_{\mu , \mu_{2} \ldots \mu_{p}}$; 
see also eq. (\ref{A-3}).)
By virtue of eq. (\ref{Maxwell-2}), the 
electromagnetic $2$-form can be expressed in terms of 
$k$, $d\phi$ and $d\psi$ as follows:
\be
F \, = \,  \frac{k}{V} \we d \phi \, + \, 
\ast (\frac{k}{V} \we d \psi) \, .
\label{def-F}
\ee
In the Appendix we show that each closed and invariant 
$2$-form gives rise to a local conservation law for a current
$1$-form (see eq. (\ref{A-7})). 
Applying this result to $F$ and $\ast F$ brings the Maxwell
equations in the form (\ref{A-8}),
\be
\cod \left( \frac{d \psi}{V} + 2 \phi \frac{\omega}{V^2} \right) 
\, = \, 0 \, , \; \;\; \; \; 
\cod \left( \frac{d \phi}{V} - 2 \psi \frac{\omega}{V^2} \right) 
\, = \, 0 \, .
\label{Maxwell-3}
\ee

As for the reduction of the Einstein equations, 
the $\Rtens(k,\, \cdot \,)$-components of the Ricci tensor are
obtained from the general
Killing field identity (\ref{A-11}) derived
in the Appendix.
Also using the expressions
$8 \pi \ast [k \we \Ttens(k)] = -2  d \phi \we d \psi$ and
$8 \pi \Ttens(k,k) = \sprod{d \phi}{d \phi} + 
\sprod{d \psi}{d \psi}$
for the electromagnetic stress energy tensor (where 
$T(k)_{\mu} \equiv T_{\mu \nu} k^{\nu}$), the general 
identities (\ref{A-12}) and (\ref{A-13}) yield
\be
\cod \left( \frac{d V}{V} \right) \, = \, 
4 \, \frac{\sprod{\omega}{\omega}}{V^2} \, - \,
2 \, \frac{\sprod{d \phi}{d \phi} + \sprod{d \psi}{d \psi}}{V} 
\label{Norm}
\ee
and
\be
d \omega \, = -2 \, d \phi \we d \psi \; \; \; 
\Longrightarrow \; \; \; \omega \, = \, dU \, + \, \psi d \phi 
\, - \, \phi d \psi \, ,
\label{Twist}
\ee
respectively, where $U$ denotes the twist potential. 
We have already argued that the Maxwell equations
for $\phi$ and $\psi$ can be cast into the form of conservation
laws (\ref{Maxwell-3}). 
This is, in fact, also true for the Poisson equation
(\ref{Norm}): Using again the identity
$\cod \left(\frac{\omega}{V^2} \right) = 0$, we have
$\cod \left( U \frac{\omega}{V^2} \right) = 
-\sprod{dU}{\frac{\omega}{V^2}}$ which, by virtue of
eq. (\ref{Twist}), brings eq. (\ref{Norm}) into the form (\ref{c-2}) 
below. We therefore end up with the following set of conserved
currents, given in terms of the four potentials
$V$, $U$, $\phi$ and $\psi$:
\bea
\cod j_N & = & \cod \left( 
\frac{\omega}{V^2} \right) \, = \, \cod \left( 
\frac{1}{V^2} (dU + \psi d \phi - \phi d \psi) \right) \, = \, 0 \, ,
\label{c-1}\\
\cod j_M & = &  \cod \left(
-\frac{1}{2} \, \frac{dV}{V} + \psi \, \frac{B}{V} + 
\phi \, \frac{E}{V} - 2U \, \frac{\omega}{V^2} \right) \, = \, 0 \, ,
\label{c-2}\\
\cod j_Q & = & \cod \left(
\frac{E}{V} - 2 \psi \, \frac{\omega}{V^2} \right) \, = \, 0 \, ,
\label{c-3}\\
\cod j_P & = &  \cod \left(
\frac{B}{V} + 2 \phi \, \frac{\omega}{V^2} \right) \, = \, 0 \, .
\label{c-4}
\eea

In addition to these equations for the electromagnetic and the
gravitational potentials,
one has the Einstein equations for the projection metric 
$\pmetric \equiv V \gtens + k \otimes k$
($\gtens$ being the spacetime metric). These are readily 
obtained from eq. (\ref{A-14}) and the fact that the 
electromagnetic stress energy tensor fulfills 
$\Ttens(X,Y) = \frac{1}{V} [ \Ttens(k,k) \gtens - \frac{1}{4 \pi} 
(d \phi \otimes d \phi + d \psi \otimes d \psi)](X,Y)$
for vector fields $X$ and $Y$ orthogonal to $k$.
The equation for the Ricci tensor of the projection metric 
$\pmetric$ thus becomes
\be
\RPtens \, = \,
\frac{1}{2V^2} \left( dV \otimes dV \right) +  
\frac{2}{V^2} \left( \omega \otimes \omega \right) -
\frac{2}{V} \left( 
d \phi \otimes d \phi + d \psi \otimes d \psi \right) \, .
\label{Ricci-proj}
\ee

It is well known -- and of crucial importance 
to what follows -- that 
the entire set of field equations 
(\ref{c-1})--(\ref{Ricci-proj})
is obtained from the effective action
(see, e.g., \cite{KSMH-80})
\be
{\cal S}_{\mbox{eff}} \, = \, \int  \Bigl( - R^{(p)} +
\frac{\sprod{dV}{dV}}{2 \, V^{2}} + 2 \, 
\frac{\sprod{\omega}{\omega}}{V^{2}} - 2 \,
\frac{\sprod{d \phi}{d \phi} + \sprod{d \psi}{d \psi}}{V}
\Bigr) \eta^{(p)} \, ,
\label{eff-1}
\ee
by considering variations with respect to the electromagnetic 
potentials $\phi$, $\psi$, the gravitational potentials $V$, $U$ and
the metric $\pmetric$ (where $\omega = \omega(U,\phi,\psi) = 
dU +\psi d \phi- \phi d \psi$). Here $R^{(p)}$ and $\eta^{(p)}$ 
denote the Ricci scalar and the volume $3$-form with respect 
to $\pmetric$. Two comments may be helpful:

First, we note that 
$\ast j = - \frac{k}{V} \we \tilde{\ast} j$
for arbitrary $1$-forms $j$ orthogonal to $k$, $\sprod{j}{k} = 0$,
(where $\tilde{\ast}$ denotes the Hodge dual with respect to 
the metric $\pmetric$).
The identity (\ref{A-5}) therefore implies that the conservation laws
$d \tilde{\ast} j = 0$ obtained from the effective action
(\ref{eff-1}) can also be written in the form $d \ast j = 0$, that is,
in the $4$-dimensional notation of eqs. (\ref{c-1})--(\ref{c-4}).

Second, the $\Rtens(k,X)$-component of the Einstein 
equations is not obtained from the effective action (\ref{eff-1})
but has already been used in order to express the
$1$-form $\omega$ in terms of the potentials $U$, $\phi$ and $\psi$.
The {\em systematic\/} 
Kaluza-Klein reduction of the Einstein-Hilbert
action in the presence of a Killing field yields an effective action
in terms of the gravitational potential $V$, the projection metric 
$\pmetric$ and the bundle connection $1$-form $\gamma$, say.
The equation for $\gamma$ then implies the existence of the 
potential $U$. Substituting $dU$ for 
$\tilde{\ast} d \gamma$ (by applying the Lagrange 
multiplier method) yields the ``partially on shell''
action (\ref{eff-1}).

\subsection{Coset Formulation}

The action (\ref{eff-1}) describes a harmonic mapping into a 
$4$-dimensional target space, effectively coupled to $3$-dimensional
gravity. Ernst \cite{E-68} was able to parametrize the target space
in terms of two complex potentials, $\Erpot$ and $\Lambda$,
\be
\Erpot \, \equiv \, V \, - \, (\phi^2 + \psi^2) \, + \,  2i \, U \, , 
\; \; \; \; \;
\Lambda \, \equiv \, - \phi \, + \, i \, \psi \, .
\label{ernst-pots}
\ee
In order to find the isometries of the target manifold, Neugebauer 
and Kramer \cite{NK-69} solved the corresponding Killing equations. 
This revealed the coset structure of the target space 
\cite{PM-83} and provided 
a parametrization of the latter in terms of the Ernst potentials
\cite{NK-69}, \cite{E-68}. (See also \cite{Ha-78}, \cite{DM-79} and
\cite{EH-79-80} for the complete integrability of the reduced system 
in the case of {\em two\/} Killing fields.) In the simplest case, 
that is for vacuum gravity, the coset space, $G/H$, is 
$SU(1,1)/U(1)$, whereas $G/H = SU(2,1)/S(U(1,1) \times U(1))$ for
the Einstein-Maxwell equations with a timelike Killing field. (If 
the dimensional reduction is performed with respect to a spacelike 
Killing field, then $H = S(U(2)\times U(1))$.)

The explicit representation of the coset manifold in terms
of the above Ernst potentials $\Erpot$ and $\Lambda$ is given by 
the hermitian matrix
\be
\Phi_{a b} \, = \, \eta_{a b} \, - \, 2 \,  \bar{v}_a v_b \, ,
\label{Kinn-1}
\ee
where $\eta = \mbox{diag} (-1,+1,+1)$, and where $v$ is the 
Kinnersley vector \cite{K-73-77}, \cite{KC-77}, 
\be
(v_0 \/ , v_1 \/ , v_2) \, = \, \frac{1}{2 \sqrt{V}} \,
(\, \Erpot-1 \, , \, \Erpot+1 \, , \, 2 \Lambda\, ) \, .
\label{Kinn-2}
\ee
It is not hard to verify that, in terms of $\Phi$, 
the effective action (\ref{eff-1}) assumes the manifestly 
$SU(2,1)$ invariant form
\be
{\cal S}_{\mbox{eff}} \, = \, \int \left( - R^{(p)} \, +
\, \mbox{Tr} \sprod{J}{J} \right) \eta^{(p)} \, ,
\; \; \; \mbox{with} \; \; 
J \, = \, \frac{1}{2} \, \Phi^{-1} d \Phi \, .
\label{eff-2}
\ee
The equations of motion following from the above action are
the $3$-dimensional Einstein equations (obtained from 
variations with respect to $\pmetric$) and the sigma-model 
equations (obtained from variations with respect to $\Phi$):
\be
\RPtens \, = \, \mbox{Tr} \{ J \otimes J \} \, ,
\; \; \; \; \; 
d \ast  J \, = \, 0 \, .
\label{Einst-sigma}
\ee
(Here we have again used the $4$-dimensional notation; see
the comment after eq. (\ref{eff-1}.)
An important feature of the coset structure is the fact that
it provides one with a set of differential equations 
which is {\em larger\/} than the original one:
Besides the $\mbox{dim}[SU(2,1)/S(U(1,1) \times U(1))] = 4$
equations (\ref{c-1})--(\ref{c-4}), the above equation
for the matrix current $J$ comprises
$\mbox{dim}[S(U(1,1) \times U(1))] = 4$ additional 
conservation laws. A straightforward (but rather unpleasant) 
computation gives the following explicit representation
for $J$:
\bdm
- J =  \left( \begin{array}{ccc}
-i j_N & j_M & 0 \\
j_M & i j_N & j_Q - i j_P \\
0 & -j_Q - i j_P & 0
\end{array} \right) + \left( \begin{array}{ccc}
i (j_1 + j_2) & i j_2 & - j_{34} \\
- i j_2 & i  (j_1 - j_2 ) & j_{34}  \\
- \overline{j_{34}}  & - \overline{j_{34}}  & -2  i j_1
\end{array} \right) \, ,
\edm
where $j_N$, $j_M$, $j_Q$ and $j_P$ were given in
eqs. (\ref{c-1})--(\ref{c-4}). The four additional currents
$j_1$, $j_2$ and $j_{34} \equiv j_3 + i j_4$ are linear 
combinations of the  $1$-forms $\omega/V^2$, $dV/V$, $E/V$ and 
$B/V$ as well. Using eqs. (\ref{c-1})--(\ref{c-4}) to 
express the latter in terms of 
$j_N$, $j_M$, $j_Q$ and $j_P$, one finds
\bea
j_1 & = & (\phi^2 + \psi^2) \, j_N \, + \,
( \psi \, j_Q \, - \, \phi \, j_P ) \, ,
\label{j-1}\\
j_2 & = & 2 \, U \, j_M \, + \, 
(\phi^2 + \psi^2 - V) \, ( \psi \, j_Q \, - \, \phi \, j_P )
\nonumber\\ & &
+ \, \frac{1}{2} \left(
1 +  4 U^2 + (\phi^2 + \psi^2 - V) \, 
(3\phi^2 + 3\psi^2 - V) \right) j_N \, , 
\label{j-2}\\
j_3 & = & \phi \, j_M \, - \, 2 \left(
- \phi U \, + \, \psi (\phi^2 + \psi^2 - \frac{V}{2}) \right) j_N
\nonumber\\ & &
- \, \frac{1}{2} \left(3 \psi^2 + \phi^2 + 1 - V) \right) j_Q
\, + \, \left( \phi \psi \, - \, U \right) j_P \, ,
\label{j-3}\\
j_4 & = & - \, \psi \, j_M \, - \, 2 \left(
\psi U \, + \, \phi (\phi^2 + \psi^2 - \frac{V}{2}) \right) j_N
\nonumber\\ & &
+ \, \frac{1}{2} \left(3 \phi^2 + \psi^2 + 1 - V) \right) j_P
\, - \, \left( \phi \psi \, + \, U \right) j_Q \, ,
\label{j-4}
\eea
It is obvious from eq. (\ref{Einst-sigma}) -- and also easy 
to verify directly from eqs. (\ref{c-1})--(\ref{c-4}) -- that 
$\cod j_{1} = \cod j_{2} = \cod j_{3} = \cod j_{4} = 0$. 
As an example,
we obtain for the first current
$\cod j_{1}$ $=$ 
$-2\sprod{\phi E + \psi B}{j_N}-\sprod{B}{j_Q}+\sprod{E}{j_P}$
$= \, 0$. (Use the identity
$\cod (f \alpha) = f \cod \alpha - \sprod{df}{\alpha}$ 
(for arbitrary
functions $f$ and $1$-forms $\alpha$) to show this.)

\subsection{Mass Formulae}

In order to apply Stokes' theorem (\ref{Stokes-1}) we 
use eq. (\ref{A-4}), which shows that each conserved current
$j$, $\cod j = 0$, gives rise to a closed $2$-form $\ast(k \we j)$,
$d\ast(k \we j)=0$. 
Using eqs. (\ref{def-F}), (\ref{c-3})
and the identity (\ref{A-5}), 
$d \left( \frac{k}{V} \right) = 2 \ast (k \we 
\frac{\omega}{V^{2}})$, 
we find, for instance,
\bdm
\ast (k \we j_{Q}) \, = \, \ast (k \we \frac{E}{V}) \, - \, 
\psi \, d \left( \frac{k}{V} \right) \, = \, \ast F
 \, - \, d \left( \psi \, \frac{k}{V} \right) \, .
\edm
In a similar way one derives the desired expressions for
$\ast (k \we j_{P})$ and $\ast (k \we j_{M})$ 
(also taking advantage of the 
identity $\ast(k \we \frac{dV}{V})$ $=$ 
$-\ast dk - 2 \frac{k}{V} \we \omega$).
The closed $2$-forms obtained from the conserved currents
(\ref{c-1})--(\ref{c-4}) become
\bea
\ast ( k \we j_N ) & = & 
\ast (k \we \frac{\omega}{V^2}) \, = \,
\frac{1}{2} \, d \left( \frac{k}{V} \right) \, , 
\label{twoform-1}\\
\ast ( k \we j_M ) &  = & \frac{1}{2} \ast dk \, + \, 
\psi \, F \, + \, \phi \ast F \, - \, 
d \left( U \,  \frac{k}{V} \right) \, ,
\label{twoform-2}\\
\ast ( k \we j_Q ) & = & \ast F \, - \, 
d \left( \psi \,  \frac{k}{V} \right) \, ,
\label{twoform-3}\\
\ast ( k \we j_P ) & = &  F \, + \, 
d \left( \phi \,  \frac{k}{V} \right) \, .
\label{twoform-4}
\eea
Stokes' theorem (\ref{Stokes-1}) now yields a set of relations 
between the charges $M$, $Q$, $P$ and the corresponding horizon 
quantities $M_{H}$, $Q_{H}$, $P_{H}$, defined by
\be
M, \, M_{H} = -\frac{1}{8 \pi} \int_{S^{2}_{\infty}, \, H} 
\ast dk \, , \; \; \; \;
Q, \, Q_{H} = -\frac{1}{4 \pi} \int_{S^{2}_{\infty}, \, H} 
\ast F \, , \; \; \; \;
P, \, P_{H} = -\frac{1}{4 \pi} \int_{S^{2}_{\infty}, \, H}  F \, ,
\label{charges-def}
\ee
where, by definition, $M_{H} = \frac{1}{4 \pi} \kappa {\cal A}$.
For asymptotically flat solutions the NUT charge vanishes and the 
integrals over the exact $2$-forms do not contribute. 
In this case, we immediately obtain from 
eqs. (\ref{twoform-2})--(\ref{twoform-4})
\be
M \, = \, M_{H} \, + \,  \phi_{H} \,  Q_{H} \, + \, \psi_{H} \, P_{H} 
\, , \; \; \; \; 
Q \, = \, Q_{H} \, , \; \; \; \; P \, = \, P_{H} \, ,
\label{mass-1}
\ee
where we have used the fact that
all potentials assume constant values on the horizon.
We also recall that asymptotic flatness implies
$V_{\infty} = 1$, whereas, by the definition of a Killing
horizon, $V_{H} = 0$.
Here and in the following we adopt a gauge for which all
other potentials vanish in the asymptotic regime,
$U_{\infty} = \phi_{\infty} = \psi_{\infty} = 0$. 
(For static, regular configurations without horizon
the above relations reduce to $M = Q = P = 0$, which yields the 
well-known non-existence theorem for self-gravitating Abelian 
{\em soliton\/} solutions \cite{BMG-88}.)

So far we have used the 
field equations to derive eqs. (\ref{mass-1}),
which imply the Smarr formula, $M = M_H + \phi_{H} Q + \psi_{H} P$.
The interesting observation is that Stokes' formula for the 
{\em additional\/} conserved currents $j_{1}$--$j_{4}$ (given in
eqs. (\ref{j-1})--(\ref{j-4})) yields a set of {\em new\/}
relations between the charges and the horizon quantities. 
Since the potentials assume constant values on the horizon, 
they can be pulled out of the integrals, which 
implies that the additional 
relations do not depend on the original ones (although, as 
already emphasized, the differential laws $\cod j_{i} = 0$ 
($i = 1 \ldots 4$) do not contain new 
information). In order to evaluate Stokes' 
theorem (\ref{Stokes-1}) for the additional four closed
$2$-forms $\ast (k \we j_i)$, one uses
$-\frac{1}{4 \pi} \int_{\infty} \ast(k \we j_{N})$ $=$
$-\frac{1}{4 \pi} \int_{H} \ast(k \we j_{N})$ $=$ $0$,
$-\frac{1}{4 \pi} \int_{\infty} \ast(k \we j_{M})$ $=$
$-\frac{1}{4 \pi} \int_{H} \ast(k \we j_{M})$ $=$ $M$,
$-\frac{1}{4 \pi} \int_{\infty} \ast(k \we j_{Q})$ $=$
$-\frac{1}{4 \pi} \int_{H} \ast(k \we j_{Q})$ $=$ $Q$ and
$-\frac{1}{4 \pi} \int_{\infty} \ast(k \we j_{P})$ $=$
$-\frac{1}{4 \pi} \int_{H} \ast(k \we j_{P})$ $=$ $P$.
In this way we immediately obtain from eqs. 
(\ref{j-1})  and (\ref{j-2}) the formulae
\be
\phi_{H} \, P \, = \, \psi_{H} \, Q \, 
\; \; \; \; \mbox{and} \; \; \;
U_{H} \, M \, = \, 0 \, ,
\label{mass-2}
\ee
respectively. Together with the Smarr formula (\ref{mass-1}), this 
enables one to solve for the horizon potentials in 
terms of the charges and $M_{H}$,
\be
\phi_{H} \, = \, Q \, \frac{M - M_{H}}{Q^{2} + P^{2}} \, , \; \; \; \; 
\psi_{H} \, = \, P \, \frac{M - M_{H}}{Q^{2} + P^{2}} \, , \; \; \; \; 
U_{H} \, = \, 0 \, ,
\label{mass-3}
\ee
where $U_{H} = 0$ reflects the fact that, for the moment, we have 
restricted ourselves to configurations with vanishing NUT charge. 
We may finally apply Stokes' theorem to either of the remaining 
eqs. (\ref{j-3}) or (\ref{j-4}). Using eq. (\ref{j-3}) we find
\be
0 \, = \, 2 \, \phi_{H} \, M \, - \, \left(
3 \psi_{H}^{2} + \phi_{H}^{2} +1 \right) Q \, + \, 
2 \left( \phi_{H} \psi_{H} - U_{H} \right) P \, .
\label{mass-4}
\ee
Substituting the expressions (\ref{mass-3}) for 
the potentials into this equation 
eventually yields the desired formula, which involves only global 
charges and the horizon quantity $M_{H}$:
\be
M^{2} \, = \, M_{H}^{2} \, + \, Q^{2} \, + \, P^{2} \, ,
\; \; \; \; \mbox{with} \; \; \; 
M_{H} \, = \, \frac{1}{4 \pi} \,  \kappa {\cal A} \, = \, 
\frac{1}{2} \, T_{H} {\cal A} \, .
\label{mass-5}
\ee
The derivation of eq. (\ref{mass-5}) implies that this 
formula holds for
every stationary, asymptotically flat black hole 
solution with 
nonrotating horizon, i.e., with Killing horizon 
generated by the 
stationary Killing field $k$. Considering the 
uniqueness theorem for the 
Reissner-Nordstr\"om metric, this is, of course, 
not surprising. However, the above derivation does, 
for instance, circumvent the 
staticity problem. Moreover, we have not required 
a nondegenerate 
horizon. Hence, the formula (\ref{mass-5}) also 
implies that the stationary, nonrotating solutions with 
vanishing surface gravity saturate the Bogomol'nyi bound 
$M^{2} = Q^{2} + P^{2}$ -- and 
vice-versa \cite{GH-82}, \cite{GHHP-83} (provided, of 
course, that 
the horizon is connected). Before we derive a similar formula 
for the EM-axion-dilaton system, we also evaluate the above 
currents for configurations with nonvanishing NUT charge.

\subsection{Mass Formulae Including NUT Charge}

The NUT charge $N$ and its horizon counterpart $N_H$ are 
defined by the boundary integrals
\be
N, \, N_{H} \, = \,
-\frac{1}{4 \pi} \int_{S^{2}_{\infty}, \, H} 
\ast (k \we j_N) \, = \,
-\frac{1}{8 \pi} \int_{S^{2}_{\infty}, \, H} 
d \left( \frac{k}{V} \right) \, .
\label{NUT-c}
\ee
Like the magnetic charge $P$, $N$ is a topological quantity.
(An illustration is provided by the Schwarzschild-NUT solution:
\be
\gtens \, = \, -V \, \left(dt - 2 N  
\cos \! \vartheta \, d \varphi \right)^{2}
\, + \, \frac{1}{V} 
\, dr^{2} \, + \, (r^{2} + N^{2}) \, d \Omega^{2} 
\, ,
\label{S-NUT}
\ee
with
\bdm
V(r) \, = \, \frac{r (r -2M) - N^{2}}{r^{2} + N^{2}} \, .
\edm
The stationary Killing $1$-form is 
$k = -V (dt - 2 N  \cos \vartheta \, d \varphi)$. Hence,
we have $d (k/V) = -2N \sin \! \vartheta d \vartheta 
\we d \varphi$ and 
$-\frac{1}{8 \pi} \int d (k/V) = N$ for any $2$-sphere;
in particular $N = N_{H}$.
Also using $\ast dk = -(r^{2} + N^{2}) 
\frac{dV}{dr} d \Omega  + (dr \we \ldots)$, one finds
\bdm
M, \, M_{H} \, = \, -\frac{1}{8\pi} 
\int_{S^{2}_{\infty}, H} \ast dk \, = \, 
\left[ \frac{2N^{2} r \, + 
\, M (r^{2} - N^{2})}{r^{2} + N^{2}} \right]_{\infty,r_{H}} \, .
\edm
As expected, the r.h.s. yields $M$ as $r \rightarrow \infty$ 
whereas, for
$r = r_{H} = M + \sqrt{M^{2} + N^{2}}$, we obtain 
$M_{H} = \sqrt{M^{2} + N^{2}}$.
For the Schwarzschild-NUT metric we therefore have
\be
N \, = \, N_{H} \, , \; \; \; \; \; 
M^{2} \, + \, N^{2} \, = \, M_{H}^{2} \, .
\label{S-NUT-2}
\ee
It will follow below, that this relation holds for any stationary,
nonrotating vacuum black hole solution.)

Let us now return to the general stationary EM equations and
evaluate Stokes' theorem for the closed $2$-forms
(\ref{twoform-1})--(\ref{twoform-4}) with nonvanishing NUT charge.
Instead of eqs. (\ref{mass-1}) we now obtain the 
slightly modified relations (in a gauge where
$\phi_{\infty} = \psi_{\infty} = U_{\infty} = 0$)
\be
N \, = \, N_H \, , \; \; \; \; 
Q \, = \, Q_H - 2 \psi_H N \, , \; \; \; \; 
P \, = \, P_H + 2 \phi_H N \, ,
\label{mass-6}
\ee
and
\be
M \, = \, M_H + \phi_H Q \, + \, \psi_H P \, - \, 2 U_H N \, .
\label{mass-7}
\ee
Here we have already used the consequence 
$\phi_H Q_H + \psi_H P_H = \phi_H Q + \psi_H P$ 
of eqs. (\ref{mass-6}) to obtain the Smarr formula 
(\ref{mass-7}) with NUT charge. 
Using the fact that the potentials assume
constant values on the horizon, we can again evaluate 
Stokes' theorem for the remaining closed 
$2$-forms $\ast(k \we j_i)$ ($i = 1 \ldots 4$), 
where now $-\frac{1}{4 \pi} \int_{\infty} \ast(k \we j_{N})$ $=$
$-\frac{1}{4 \pi} \int_{H} \ast(k \we j_{N})$ $=$ $N$. 
The expressions
(\ref{j-1})--(\ref{j-4}) then imply the following relations
(with $V_{\infty} = 1$ and $V_H = 0$)
\bea
0 & = &
(\phi_{H}^2 + \psi_{H}^2) \, N \, + \, (\psi_{H} Q - \phi_{H} P) \, ,
\nonumber \\
N & = &
2 U_H M + (\phi_{H}^2 + \psi_{H}^2) (\psi_{H} Q - \phi_{H} P) + 
\frac{1}{2} \left[ 1 + 4 U_H^2 + 
3 (\phi_{H}^2 + \psi_{H}^2)^2 \right] N \, ,
\nonumber \\
0 & = &
\phi_{H} M + 2 \left[\phi_{H} U_H - \psi_{H} (\phi_{H}^2 + \psi_{H}^2)\right] N
- \frac{1}{2} (3 \psi_{H}^2 + \phi_{H}^2 + 1) Q
+ (\phi_{H} \psi_{H} - U_H) P \, ,
\nonumber \\
0 & = &
\psi_{H} M + 2 \left[\psi_{H} U_H + \phi_{H} (\phi_{H}^2 + \psi_{H}^2)\right] N
- \frac{1}{2} (3 \phi_{H}^2 + \psi_{H}^2 + 1) P
+ (\phi_{H} \psi_{H} + U_H) Q \, .
\nonumber
\eea
Adding together $\phi_{H}$ times the third and $\psi_{H}$ times the 
fourth relation and using the Smarr formula (\ref{mass-7}) one 
finds $(\phi_{H}^2 + \psi_{H}^2)(M + M_H) = \phi_{H} Q + \psi_{H} P$.
In combination with the first of the above formulae, this enables 
one to solve for $\phi_{H}$ and $\psi_{H}$. Substituting 
the result into the Smarr formula then also yields $U_H$:
\be
\phi_H \, = \, 
\frac{(M + M_H) Q - N P}{(M + M_H)^2 + N^2} \, , \; \; \; \; \;
\psi_H \, = \, 
\frac{(M + M_H) P + N Q}{(M + M_H)^2 + N^2} \, ,
\label{phipsi}
\ee
\be
U_H \, = \, 
\frac{(M + M_H) \, (Q^2 + P^2 + M_H^2 - M^2) - 
(M - M_H) \, N^2}{ 2N \, [(M + M_H)^2 \, + \, N^2 ]} \, ,
\label{uh}
\ee
We may finally use the expressions for the horizon 
potentials in the second of the above formulae, which can also be 
written in the form $4 U_H N (M + U_H N) = N^2 - 
(\phi_{H} P - \psi_{H} Q)^2$.
A short computation yields the desired relation between the total 
charges and the horizon quantity 
$M_H = \frac{1}{4 \pi} \kappa {\cal A}$,
\be
M^2 \, + \, N^2 \, = \, 
\left(\frac{1}{4 \pi} \kappa {\cal A}\right)^2 \, + \, 
Q^2 \, + \, P^2 \, ,
\label{res-end-1} 
\ee
which generalizes the previous result (\ref{mass-5}).

\section{The Einstein-Maxwell-Axion-Dilaton System}

Let us now consider the bosonic sector of $4$-dimensional 
heterotic string theory or, equivalently, $N=4$ supergravity
with one vector field. Denoting the dilaton scalar field with $S$,
the axion pseudoscalar field with $\kappa$ and the Abelian (Maxwell) 
vector field with $A$, the effective action can be cast 
into the form
\be
{\cal S} \, = \, \frac{1}{16 \pi} \int \left[
- \ast \! R \, + \, 2 F \we \ast G \, + \, 
2 \, dS \we \ast dS \, + \, \frac{1}{2}e^{4S} \, d \kappa \we
\ast d \kappa \right] \, ,
\label{dil-ax-act}
\ee
where $F$ is the field strength of the vector field.
Here we have introduced the $2$-form $G$, which turns out to be
very convenient in the following. For vanishing dilaton and
axion fields we have $G \rightarrow F$, whereas, in general,
$G$ is a combination of $F$ and $\ast F$, involving the dilaton 
and the axion fields:
\be
G \, \equiv \, e^{-2S} F \, - \, \kappa \ast F \, , \; \; \; 
\mbox{where} \; \; F \, = \, dA \, .
\label{def-G}
\ee
(Hence, $F \we \ast G = e^{-2S} F \we \ast F + \kappa F \we F$.)
It is also worthwile recalling that
it is the boundary integral over $\ast G$ 
(rather than $\ast F$) which is identified with the electric charge 
in the presence of a dilaton and an axion 
(see, e.g., \cite{GKK-96} and eq. (\ref{QPDA-1}) \\below).

\subsection{Dimensional Reduction}

The dimensional reduction of the field equations in the presence of 
the stationary Killing field $k$ can be performed along the same
lines as for the EM system discussed in the previous section.
The Bianchi identity, $dF = 0$, and the ``Maxwell'' equation, 
$d \ast G = 0$ (i.e., the variational equation with respect to $A$), 
give (locally) again rise to two scalar potentials, 
$\phi$ and $\psi$, say:
Since $L_k F = 0$ and $L_k \ast G = \ast L_k G = 0$ one 
obtains (with $L_k = i_K \circ d + d \circ i_k$) the equations 
$d (i_k F) = 0$ and $d (i_k \ast G) = 0$, and therefore
\be
d \phi \, = \, -i_k F \, , \; \; \; \; \;
d \psi \, = \, i_k \ast G \, .
\label{bas-3bis}
\ee
Since both $F$ and $\ast G$ are closed and invariant with respect to
the Killing field $k$, we can apply the construction discussed in
the Appendix (see eq. (\ref{A-7})) to obtain two conserved current 
$1$-forms,
\bea
\cod j_{P} & = & 0 \, , \; \; \; \mbox{where} \; \; 
j_{P} \, = \, \frac{\BB}{V} \, + \, 2 \phi \,
\frac{\omega}{V^{2}} \, , \; \; \; \BB \equiv i_{k} \ast F \, ,
\label{bas-4-0}\\
\cod j_{Q} & = & 0 \, , \; \; \; \mbox{where} \; \;  
j_{Q} \, = \, \frac{\EE}{V} \, - \, 2 \psi \,
\frac{\omega}{V^{2}} \, , \; \; \; \EE \equiv - i_{k} G \, .
\label{bas-4}
\eea
It is easy to see that the $1$-forms $\EE$ and $\BB$ are 
linear combinations in $d \phi$ and $d \psi$,
\be
\left(\begin{array}{c}
\EE \\ \BB
\end{array}\right) 
\, = \, 
{\cal D} \, 
\left(\begin{array}{c}
d \phi \\ d \psi
\end{array}\right) \, , \; \; \; \mbox{with} \; \; 
{\cal D} \, = \, 
\left(\begin{array}{cc}
e^{-2S} + \kappa^2 \/ e^{2S} & \kappa \, e^{2S} \\
\kappa \, e^{2S} & e^{2S}
\end{array}\right) \, .
\label{matrix-EB}
\ee
In terms of $\EE$, $\BB$, the potentials
and the Killing $1$-form we also have
\be
F \, = \, \frac{k}{V} \we d \phi \, + \, 
\ast (\frac{k}{V} \we \BB) \, , \; \; \; \; \; 
G \, = \,  \frac{k}{V} \we \EE \, + \, 
\ast (\frac{k}{V} \we d \psi) \, ,
\label{FG}
\ee
which generalizes eq. (\ref{def-F}). 
It is worth recalling that the
symmetric and simplectic matrix ${\cal D}$ is a special case of 
the matrix introduced in \cite{BMG-88}, parametrizing an 
arbitrary number of moduli fields (see also \cite{GKK-96}).
(For vanishing axion and dilaton fields we have 
$G \rightarrow F$, ${\cal D} \rightarrow \idid$, 
$\EE \rightarrow d \phi$ and
$\BB \rightarrow d \psi$, which shows that eqs. (\ref{bas-4-0}) and
(\ref{bas-4}) reduce to the ordinary Maxwell 
equations (\ref{Maxwell-3}) in this case.)

The axion and dilaton equations are obtained from
variations of the action (\ref{dil-ax-act}) 
with respect to $\kappa$ and
$S$, respectively. One finds
\be
\cod \left( e^{4S} d \kappa \right) \, = 
\,- 2 \sprod{F}{\ast F} \, ,
\label{bas-5}
\ee
\be
\cod \left( dS \, - \, \frac{1}{2} e^{4S} \kappa \, d \kappa
\right) \, = \, \sprod{F}{G} \, .
\label{bas-6}
\ee
(Note that the variation with respect to the dilaton field first
gives 
$\cod dS + \frac{1}{2} e^{4S} \sprod{\kappa}{\kappa} = 
e^{-2S} \sprod{F}{F}$. Integrating by parts and using the
axion equation (\ref{bas-5}) and the definition of $G$ then 
yields eq. (\ref{bas-6}).
Also note that 
$\sprod{\alpha}{\beta} \ast 1 \equiv \alpha \we \ast \beta$
for arbitrary forms of the same degree; hence
$\sprod{F}{G} = \frac{1}{2} F_{\mu \nu} G^{\mu \nu}$.)
Now using the ``Maxwell'' equations
(\ref{bas-4-0}) and (\ref{bas-4}), the identity 
$\cod \left( \frac{\omega}{V^2} \right)=0$ and the formula
$\cod (f \alpha) = f \cod \alpha - \sprod{df}{\alpha}$ 
(for arbitrary 
functions $f$ and $1$-forms $\alpha$), we can write
the axion and dilaton equations 
(\ref{bas-5}) and (\ref{bas-6}) in the form of
current conservation laws as well, since
\bdm
\sprod{F}{\ast F} =
-2 \, \cod \left( \phi \frac{\BB}{V} + \phi^2 \frac{\omega}{V^2}
\right) \, , \; \; \; \; 
\sprod{F}{G} = 
\cod \left( \phi \frac{\EE}{V} - \psi \frac{\BB}{V} 
-2 \phi \psi \frac{\omega}{V^2} \right) \, .
\edm 

It remains to consider the Einstein equations. In order to 
evaluate the Poisson equation (\ref{A-12}) and the
twist equation (\ref{A-13}),
we have to compute the Ricci $1$-form $\Rtens(k)$. 
Since the kinetic terms of the
axion and the dilaton do not contribute to $\Rtens(k)$, we have
$R(k)_{\mu} = [ t_{\mu \nu} - \frac{1}{2} g_{\mu \nu}
t^{\sigma}_{\; \sigma} ] k^{\nu}$, where $t_{\mu \nu}$
is the stress-energy tensor of the vector field,
\be
t_{\mu \nu} \, = \, \frac{1}{8 \pi} \left[ 2 \, F_{\mu \sigma} 
G^{\sigma}_{\; \nu} - g_{\mu \nu} \sprod{F}{G} \right] \, .
\label{bas-7}
\ee
Contracting with $k^{\nu}$ and using the expressions (\ref{FG}),
eqs. (\ref{A-12}) and (\ref{A-13}) yield
\be
\cod \left( \frac{d V}{V} \right) \, = \, 
\frac{4}{V^2} \sprod{\omega}{\omega} \, - \, 
\frac{2}{V} \, 
\left( 2 \sprod{i_kF}{i_kG} + V \sprod{F}{G} \right) \, ,
\label{bas-9} 
\ee
and
\be
d \omega = -2 \, d \phi \we d \psi \, ,
\label{bas-8}
\ee
respectively.
The twist equation (\ref{bas-8}) implies the 
existence of a twist potential $U$, such that 
$dU = \omega + \phi d \psi - \psi d \phi$. The Poisson
equation (\ref{bas-9}) therefore also assumes the form 
of a conservation law, since its r.h.s. becomes
\bdm
\frac{4}{V^2} \sprod{\omega}{\omega} - \frac{2}{V} \left(
\sprod{d \phi}{\EE} + \sprod{d \psi}{\BB} \right) =
\cod \left( 2 \phi \frac{\EE}{V} + 2 \psi \frac{\BB}{V} - 
4 U \frac{\omega}{V^2} \right) \, .
\edm

In conclusion, the field equations for the three pairs of scalar 
potentials $(V,U)$, $(\phi, \psi)$ and $(S,\kappa)$ 
can be cast into the form of
six conservation laws for the following current $1$-forms (see
\cite{BMG-88} for the more general case of an arbitrary number of 
moduli fields):
\bea
\cod j_N & = & \cod \left( \frac{\omega}{V^2} \right) = 0 \, 
\; \; \; \mbox{where} \; \; 
\omega = dU + \psi d \phi - \phi d \psi \, ,
\label{JJN} \\
\cod j_M & = & \cod \left(-\frac{1}{2} \, \frac{dV}{V} + 
\psi \, \frac{\BB}{V} + \phi \, 
\frac{\EE}{V} - 2U \, \frac{\omega}{V^2} 
\right) = 0 \, ,
\label{JJM} \\
\cod j_Q & = & \cod 
\left( \frac{\EE}{V} - 2 \psi \, \frac{\omega}{V^2} 
\right) = 0 \, ,
\label{JJQ} \\
\cod j_P & = & \cod 
\left( \frac{\BB}{V} + 2 \phi \, \frac{\omega}{V^2} 
\right) = 0 \, ,
\label{JJP} \\
\cod j_A & = & \cod \left( e^{4S} \, d \kappa +
4 \phi \, \frac{\BB}{V} + 4 \phi^2 \, 
\frac{\omega}{V^2} \right) = 0 \, ,
\label{JJA} \\
\cod j_D & = & \cod \left( -2 
\, dS + e^{4S} \kappa \, d \kappa -
2 \psi \, \frac{\BB}{V} + 2 \phi 
\, \frac{\EE}{V}  - 4 \phi \psi \,
\frac{\omega}{V^2} 
\right) = 0 \, ,
\label{JJD}
\eea
where $\EE$ and $\BB$ are defined in terms of $S$, $\kappa$, 
$d \phi$ and $d \psi$ by 
eq. (\ref{matrix-EB}). It may be worth
noticing that eqs. (\ref{JJN})--(\ref{JJP}) 
reduce to the corresponding
Einstein and Maxwell 
equations (\ref{c-1})--(\ref{c-4}) for vanishing 
dilaton and axion fields. However, for $\kappa = S = 0$, 
the {\em entire} set of equations
(\ref{JJN})--(\ref{JJD}) is not equivalent to eqs. 
(\ref{c-1})--(\ref{c-4}), since the dilaton and axion equations 
(\ref{JJA}) and (\ref{JJD}) impose additional 
restrictions to the vector 
field $A$. It is for this reason that the coset 
formulation to be
discussed below does not reduce to the electrovac coset 
representation for $\kappa = S = 0$.

The remaining equations, which will not be used in the 
following, are the Einstein equations for the projection metric 
$\pmetric = V \gtens + k \otimes k$.
These are again obtained from 
the reduction formula (\ref{A-14}), using
$R_{\mu \nu} = 8 \pi t_{\mu \nu} + 2 S_{\mu} S_{\nu} + 
\frac{1}{2} e^{4S} \kappa_{\mu} \kappa_{\nu}$. Also 
taking advantage of the expression (\ref{bas-7}) for 
$t_{\mu \nu}$, the Einstein equations for $\pmetric$ become
\bea
\RPtens & = &
\frac{1}{2 V^2} \left( dV \otimes dV \right)
 \, + \, \frac{2}{V^2} \left( \omega \otimes \omega \right)
\, - \, \frac{2}{V} \, \left(
d \phi \otimes \EE + d \psi \otimes \BB \right) \nonumber \\
&  & + \, 2 \, dS \otimes dS \, + \, \frac{1}{2} e^{4S} \,
d \kappa \otimes d \kappa \, ,
\label{bas-10}
\eea
which reduces to eq. (\ref{Ricci-proj}) for the EM case.

\subsection{Coset Representation}

The entire set of field equations, i.e., the conservation laws
(\ref{JJN})--(\ref{JJD}) and the $3$-di\-men\-sion\-al
Einstein equations (\ref{bas-10}) can be 
obtained from variations of the effective action,
${\cal S}_{\mbox{eff}}$,
with respect to the scalar fields $V$, $U$, $\phi$, $\psi$,
$S$, $\kappa$ and the projection metric
$\pmetric$, where
\bea
{\cal S}_{\mbox{eff}} & = & \int  \Bigl[ - R^{(p)} +
\frac{\sprod{dV}{dV}}{2 \, V^{2}} + 2 \, 
\frac{\sprod{\omega}{\omega}}{V^{2}} - 2 \,
\frac{\sprod{d \phi}{\EE} + \sprod{d \psi}{ \BB}}{V}
\nonumber\\ & &
+ \, 2 \, \sprod{dS}{dS} \, + \, \frac{1}{2} \,
e^{4S} \, \sprod{d \kappa}{d \kappa}
\Bigr] \, \eta^{(p)} \, .
\label{eff-act-ADMAX}
\eea
(Recall that $\omega = dU - \phi d \psi + \psi d \phi$,
and $\EE$ and $\BB$ are given in terms of the potentials by
eq. (\ref{matrix-EB}).)
Combining the Maxwell 
potentials into a vector, $\ul{a} \equiv (\phi, \psi)^{T}$, 
and using the matrix ${\cal D}$ defined in eq. 
(\ref{matrix-EB}), the effective action assumes the compact form
\be
{\cal S}_{\mbox{eff}} =  \int  \Bigl[ - R^{(p)} +
\frac{\sprod{dV}{dV}}{2  V^{2}} + 2  
\frac{\sprod{\omega}{\omega}}{V^{2}} - 2 
\frac{\sprod{d\ul{a}^T}{{\cal D} d\ul{a}}}{V} +
\sprod{{\cal D}^{-1} d{\cal D}}{{\cal D}^{-1}d{\cal D}} \Big]
\, \eta^{(p)} \, ,
\label{eff-act-2}
\ee
where the inner product in the last two terms also involves the 
matrix trace.
In terms of $\ul{a}$ and the 
antisymmetric $2\times 2$ tensor $\varepsilon$ 
one has $\omega = \omega(U,\ul{a}) = 
dU + \ul{a}^T \varepsilon^{-1} d\ul{a}$.
The equations (\ref{JJN})--(\ref{JJP}) 
are obtained from variations with 
respect to the gravitational potentials 
$V$ and $U$ and the potential 
$\ul{a}$. Since ${\cal D}$ is 
symmetric and symplectic, the axion and 
the dilaton describe a non-linear sigma-model with coset space
$\bar{G}/\bar{H}$ $=$ $SL(2,\RR) / U(1)$ 
(see, e.g., \cite{DG-96-Square}).
Hence, the variation of the above action 
with respect to ${\cal D}$ yields the axion and 
dilaton equations
(\ref{JJA}) and (\ref{JJD})
{\em and\/} an additional equation.
In fact, one easely finds the following additional conserved 
current:
\be
\cod j_{AD} \, \equiv \, \cod \left(
4 \kappa \, dS + (1 - \kappa^2 e^{4S}) \, d \kappa + 
4 \psi \, \frac{\EE}{V} - 4 \psi^2 \, \frac{\omega}{V^2}
\right) \, = \, 0 \, .
\label{JJAD}
\ee
This formula is, of course, a consequence of the set of 
field equations (\ref{JJN})--(\ref{JJD}), as 
can also be verified directly. However, its 
integrated version will provide us with an additional 
relation between the charges and the horizon quantities. 
The axion and dilaton equations 
(\ref{JJA}) (\ref{JJD}) and (\ref{JJAD}) assume the form 
$\cod{\cal J} = 0$, where
\be
{\cal J} \, = \, {\cal D}^{-1} d {\cal D} \, + \, 
4 \, (\ul{a}^T \otimes \varepsilon \ul{a}) \, j_N \, + \, 
2 \, [ \ul{a} \otimes \ul{j} - 
\varepsilon^{-1} (\ul{a} \otimes \ul{j})^T \varepsilon ] \, ,
\label{JJJ}
\ee 
and where we have introduced the notations
\be
{\cal J} \, \equiv \,  
\left( \begin{array}{cc}
j_D & j_A \\ j_{AD} & - j_D 
\end{array} \right) \, , \; \; \; \; \; 
\ul{j} \, \equiv \,  
\left( \begin{array}{c}
j_Q \\ j_P \end{array} \right) \, .
\label{def-currents}
\ee
(In deriving eq. (\ref{JJJ}) we have also used 
eqs. (\ref{JJN}), (\ref{JJQ}) and (\ref{JJP}) to substitute
the $1$-forms $\frac{\omega}{V^2}$, $\frac{\EE}{V}$ and 
$\frac{\BB}{V}$ by the currents $j_N$, $j_Q$ and $j_P$.)
Before we proceed, we recall some facts concerning the structure
of the stationary EMDA system.

Since $SL(2,\RR)$ is isomorphic to $SU(1,1)$, 
the axion and dilaton 
describe a nonlinear sigma-model with the same coset space
as the vacuum Ernst system, $SU(1,1)/SU(1)$. 
(In fact, using the complex target space
coordinate $z = \kappa + i e^{-2S}$, the effective density
$\mbox{Tr} \sprod{{\cal D}^{-1} 
d {\cal D}}{{\cal D}^{-1} d {\cal D}}$
becomes $\sprod{dz}{d \bar{z}}/(z-\bar{z})^2$, 
which is the same 
expression as one finds for the vacuum Ernst potential. For
axion-dilaton gravity without vector fields, the K\"ahler 
metric on the target space is therefore generated by the 
potential $\ln (V \, e^{-2S})$; 
see \cite{DG-96-Square} for details). 

The action (\ref{eff-act-ADMAX}) obviously describes a 
harmonic mapping which is
effectively coupled to $3$-dimensional gravity. 
This is indeed the case
for an arbitrary number of self-gravitating Abelian vector 
fields coupled
to massless scalar (moduli) fields which form a coset space 
$\bar{G} / \bar{H}$. Breitenlohner {\it et al.} 
\cite{BMG-88} have given a
classification of models which admit a sufficiently 
large symmetry group, such that the {\em entire\/} set 
of potentials -- i.e., 
the moduli {\em and\/} the vector and gravitational 
potentials -- form a coset space, $G/H$. 

The $SL(2,\RR)$ axion-dilaton symmetry is still present in
axion-dilaton gravity with an Abelian gauge field. 
Like in the EM case,
the system also possesses an $SO(1,2)$ symmetry, arising 
from the dimensional 
reduction with respect to the Abelian isometry group 
generated by the
Killing field. Gal'tsov and Kechkin \cite{GK-94} have shown 
that the full
symmetry group is, however, larger than 
$SL(2,\RR) \times SO(1,2)$.
Indeed, the target space for dilaton-axion gravity 
with an $U(1)$ vector field is the coset 
$SO(2,3)/(SO(2) \times SO(1,2))$ \cite{DG-95}. 
Using the fact that $SO(2,3)$ is isomorphic to $Sp(4,\RR)$, 
Gal'tsov and 
Kechkin \cite{GK-95} were also able to give a 
parametrization of the 
target space in terms of $4 \times 4$ (rather than 
$5 \times 5$) matrices. The relevant coset was shown to be 
$Sp(4,\RR)/U(1,1)$, which implies that, besides the field 
equations (\ref{JJN})--(\ref{JJD}), there exists a set of 
{\em four\/} additional conserved currents (one of which,
$j_{AD}$, was already constructed
above from the $SL(2,\RR)$ symmetry).

The explicit representation of the target space in terms of the 
potentials ($V$, $U$), ($\phi$, $\psi$) and ($S$, $\kappa$) 
is given by the symplectic $4 \times 4$ matrix $\Phi$,
\be
\Mtens \, = \,
\left( \begin{array}{cc}
\Ptens^{-1} & \Ptens^{-1} \Qtens \\
\Qtens \Ptens^{-1} & \Ptens + \Qtens \Ptens^{-1} \Qtens
\end{array} \right) \, ,
\label{M-matr}
\ee
where $\Ptens$ and $\Qtens$ are the $2 \times 2$ matrices
\be
\Ptens = e^{-2S}
\left( 
\begin{array}{cc}
e^{2S} V - 2 \phi^2 & \sqrt{2} \phi \\
\sqrt{2} \phi & -1
\end{array}
\right) , \; \; \; 
\Qtens = 
\left( 
\begin{array}{cc}
-2 \phi ( \psi + \kappa \phi) - 2 U & \sqrt{2} (\psi + \kappa \phi) \\
\sqrt{2} (\psi + \kappa \phi) & - \kappa
\end{array}
\right) \, ;
\label{Q-matr}
\ee
see, e.g., \cite{G-96}, \cite{CG-96}. 
(Our potentials slightly differ 
form the ones used in \cite{CG-96}: The potential pairs 
$(f,\chi)$, $(v,u)$ and $(\kappa,\phi)$ of \cite{CG-96} 
are our $(V,U)$, 
$(-\sqrt{2}\phi, \sqrt{2}\psi)$ and $(\kappa,S)$, respectively.)
In trems of the matrix $\Mtens$ the 
effective action (\ref{eff-act-2}) 
assumes the desired form
\be
{\cal S}_{\mbox{eff}} \, = \,  \int  \left[ - \, R^{(p)} \, + \,
\mbox{Tr} \, 
\sprod{\Mtens^{-1} \dMtens}{\Mtens^{-1} \dMtens} \, \right] \,
\eta^{(p)} \, ,
\label{eff-act-3}
\ee
where the trace-free matrix $\Mtens^{-1} d\Mtens$ comprises four
$2 \times 2$ current matrices, three of which are algebraically 
independent. A lengthy computation yields the following explicit 
expressions for the latter in terms of the ten currents
$j_N$, $j_M$, $j_Q$, $j_P$, $j_A$, $j_D$, $j_{AD}$ and 
$j_{1}$--$j_{3}$:
\be
\Ptens^{-1} \dQtens \Ptens^{-1} \, = \, -
\left( 
\begin{array}{cc}
2 \,  j_N & \sqrt{2} \, j_P \\
\sqrt{2} \, j_P &  j_A
\end{array}
\right) \, ,
\label{curr-1-matr}
\ee
\be
\Qtens \Ptens^{-1} \dQtens \Ptens^{-1} + \dPtens \Ptens^{-1} \, =
\left( 
\begin{array}{cc}
- 2 \,  j_M & - \sqrt{2} \, j_1 \\
\sqrt{2} \, j_Q &  j_D
\end{array}
\right) \, ,
\label{curr-2-matr}
\ee
\be
\dQtens - \dPtens \Ptens^{-1} \Qtens - \Qtens \Ptens^{-1} \dPtens
- \Qtens \Ptens^{-1} \dQtens \Ptens^{-1} \Qtens =
\left( 
\begin{array}{cc}
-2 \,  j_3 & \sqrt{2} \, j_2 \\
\sqrt{2} \, j_2 & -  j_{AD}
\end{array}
\right) \, .
\label{curr-3-matr}
\ee
The conservation laws for the currents 
$j_N$, $j_M$, $j_Q$, $j_P$, $j_A$, $j_D$ are identical with the
field equations (\ref{JJN})--(\ref{JJD}). The conserved 
current $j_{AD}$, arising from the dilaton-axion symmetry,
was given in eq. (\ref{JJAD}). The remaining additional 
conserved currents, 
$j_1$--$j_3$, can be expressed in terms of $j_N$, $j_M$, 
$\ul{j}=(j_Q, j_P)^T$ and the $2 \times 2$ matrix 
${\cal D}^{-1} d {\cal D}$ as follows:
\be
\left( \begin{array}{c} j_1 \\ j_2
\end{array} \right) \, = \, \left(
{\cal D}^{-1}d{\cal D} \, + \, 2 \idid j_M \right) \ul{a} \, + \, 
\left( V {\cal D}^{-1} \, - \, 2 \/ U \/ \varepsilon \right) \,
\left( \ul{j} + 2 \, j_N \,  \varepsilon \ul{a} \right) \, , 
\label{j1j2}
\ee
\be
j_3 \, = \, \mbox{Tr} \Bigl\{
\ul{a}^T \varepsilon \left[
({\cal D}^{-1} d{\cal D}) \ul{a} \, + \, 
2 V \, {\cal D}^{-1} \left( \ul{j} + 2 \, j_N \, 
\varepsilon \ul{a} \right)
\right] \Bigr\}  
\, + \, 4 U \, j_M \, + \, (V^2 + 4 U^2) \, j_N \, .
\label{j3}
\ee

\subsection{Mass Formulae}

In order to apply Stokes' theorem (\ref{Stokes-1}), we 
have to compute the closed $2$-forms $\ast(k \we j)$ 
obtained from the ten conserved 
currents $j_N$, $j_M$, $\ul{j}$, ${\cal J}$, $(j_1,j_2)$ and $j_3$, 
given in eqs. (\ref{JJN})--(\ref{JJP}), (\ref{JJJ}), (\ref{j1j2}) 
and (\ref{j3}), respectively (see eq. (\ref{def-currents}) for the 
definitions of $\ul{j}$ and ${\cal J}$). 
To this end, we first 
express the $2$-forms arising from the 
gravitational and the electromagnetic 
currents (\ref{JJN})--(\ref{JJP}) 
in terms of the $2$-forms $\ast dk$, $d (\frac{k}{V})$, 
$\ast G$ and $F$
(which give rise to the mass, the NUT charge and the 
electric and magnetic charges, respectively). 
This is achieved in a similar way as in the EM case. One finds
\bea
\ast ( k \we j_N ) & = & 
\ast (k \we \frac{\omega}{V^2}) \, = \,
\frac{1}{2} \, d \left( \frac{k}{V} \right) \, , 
\label{twoform-1-n}\\
\ast ( k \we j_M ) &  = & \frac{1}{2} \ast dk \, + \, 
\psi \, F \, + \, \phi \ast G \, - \, 
d \left( U \,  \frac{k}{V} \right) \, ,
\label{twoform-2-n}\\
\ast ( k \we j_Q ) & = & \ast G \, - \, 
d \left( \psi \,  \frac{k}{V} \right) \, ,
\label{twoform-3-n}\\
\ast ( k \we j_P ) & = &  F \, + \, 
d \left( \phi \,  \frac{k}{V} \right) \, .
\label{twoform-4-n}
\eea
(For vanishing axion and dilaton fields this reduces to the
corresponding EM expressions (\ref{twoform-1})--(\ref{twoform-4}),
since then $G \rightarrow F$.)
The following integrals over $S^2_{\infty}$ and
$H = \Sigma \cap {\cal H}$ give the electric, 
magnetic, dilaton and axion charges, 
and their counterparts, $Q_H$, $P_H$, $D_H$ and $A_H$ defined on the
horizon:
\bea
Q, \, Q_{H} = -\frac{1}{4 \pi} \int_{S^{2}_{\infty}, \, H} 
\ast G \, ,
& & P, \, P_{H} = -\frac{1}{4 \pi} \int_{S^{2}_{\infty}, \, H}  F \, ,
\label{QPDA-1} \\
D, \, D_{H} = - \frac{1}{8 \pi} \int_{S^{2}_{\infty}, \, H} 
\ast (k \we j_{D}) \, , 
& & A, \, A_{H} = -\frac{1}{8 \pi} \int_{S^{2}_{\infty}, \, H}  
\ast (k \we j_{A}) \, .
\label{QPDA}
\eea
Requiring that both $S$ and $\kappa$ remain finite on the horizon, 
we find from the general properties of Killing horizons that
\be
D_H \, = \, 0 \, , \; \; \; \; \; \; 
A_H \, = \, 0 \, .
\label{DHAH}
\ee
We recall that the total mass 
$M$ and the corresponding horizon quantity 
$M_H = \frac{1}{4 \pi} \kappa {\cal A}$ are given by the
Komar integrals over $S^2_{\infty}$ and $H$. In a similar way
one obtains the NUT charge $N$ and its horizon counterpart $N_H$:
\be
M, \, M_{H} = -\frac{1}{8 \pi} \int_{S^{2}_{\infty}, \, H} 
\ast dk \, , \; \; \; \; \; 
N, \, N_{H} = -\frac{1}{8 \pi} \int_{S^{2}_{\infty}, \, H}
d \left( \frac{k}{V} \right) \, .
\label{MN}
\ee

We may now apply Stokes' theorem (\ref{Stokes-1}) to the closed
$2$-forms (\ref{twoform-1-n})--(\ref{twoform-4-n}). 
Adopting a gauge for which the electromagnetic and the twist  
potentials vanish at infinity, $\phi_{\infty} = 
\psi_{\infty} = U_{\infty} = 0$, 
we immediately obtain the relations
\be
N \, = \, N_H \, , \; \; \; \; 
Q \, = \, Q_H - 2 \psi_H N \, , \; \; \; \; 
P \, = \, P_H + 2 \phi_H N \, ,
\label{rel-1}
\ee
and
\be
M \, = \, M_H + \phi_H Q \, + \, \psi_H P \, - \, 2 U_H N \, ,
\label{rel-2}
\ee
where we have already used eqs. (\ref{rel-1}) on the r.h.s.
of the Smarr formula (\ref{rel-2}), i.e., we have replaced
$\phi_H Q_H + \psi_H P_H$ by $\phi_H Q + \psi_H P$.

The information from the
remaining conservation laws is now extracted as follows:
First, we choose $S_{\infty} = \kappa_{\infty} = 0$.
(This can by achieved by generalized Ehlers and 
Harrison transformations; see \cite{GL-EH}). 
The currents $j_{A}$ and $j_{AD}$ then coincide at
infinity and the definitions (\ref{QPDA}) of the dilaton and axion
charges yield
(with $\ul{a}_{\infty} = (\phi_{\infty}, \psi_{\infty})^{T} = 0$)
\be
-\frac{1}{4 \pi} 
\, \int_{S^{2}_{\infty}} \ast (k \we {\cal D}^{-1}d{\cal D}) 
\, = \, 2 \left( \begin{array}{cc}
D & A \\ A & - D \end{array} \right) \, , \; \; \; \; \;  
\int_{H} \ast (k \we {\cal D}^{-1}d{\cal D}) \, = \, 0 \, .
\label{d-1dd}
\ee
(In the second integral we have used eq. (\ref{DHAH}) and
the fact that $\kappa$ and $S$
assume constant values on the horizon.)
Since all potentials can be
pulled out of the integrals, we are now able to evaluate
Stokes' theorem for the remaining closed $2$-forms;
$\ast(k \we {\cal J})$ and
$\ast(k \we j_{i}$ ($i = 1 \ldots 3$), using
$-\frac{1}{4 \pi} \int_{\infty} \ast(k \we j_{N})$ $=$
$-\frac{1}{4 \pi} \int_{H} \ast(k \we j_{N})$ $=$ $N$,
$-\frac{1}{4 \pi} \int_{\infty} \ast(k \we j_{M})$ $=$
$-\frac{1}{4 \pi} \int_{H} \ast(k \we j_{M})$ $=$ $M$ and
$-\frac{1}{4 \pi} \int_{\infty} \ast(k \we \ul{j})$ $=$
$-\frac{1}{4 \pi} \int_{H} \ast(k \we \ul{j})$ $=$ $(Q,P)$.

From the closed matrix $2$-form $\ast(k \we {\cal J})$
(with ${\cal J}$ given in eq. (\ref{JJJ})) 
we obtain an expressions for the dilaton
charge $D$ and two expressions for the axion charge $A$.
Combining these  gives
\bea
D & = & \phi_H \, (Q + N \psi_H) \, - \, 
        \psi_H \, (P - N \phi_H) \, ,
\nonumber\\
A & = & \psi_H \, (Q + N \psi_H) \, + \, 
        \phi_H \, (P - N \phi_H) 
\label{rel-3}
\eea
and
\be
N \, (\phi_H^{\, 2} + \psi_H^{\, 2}) \, = \, 
\phi_H P - \psi_H Q \, .
\label{rel-4}
\ee
Stokes' theorem for $\ast(k \we j_3)$ (with $j_3$ given in 
eq. (\ref{j3})) is easely evaluated, since the trace term gives no 
contribution at the horizon ($V_H = 0$, 
$({\cal D}^{-1}d{\cal D})_H = 0$) and also vanishes at
infinity ($\ul{a}_{\infty} = 0$). We thus have
from eq. (\ref{j3})
\be
N \, = \, 4 \, U_H \, (M + U_H N) \, .
\label{rel-6}
\ee
Finally, the evaluation of the $2$-forms
$\ast(k \we j_{1})$ and $\ast(k \we j_{2})$ (with $j_1$ 
and $j_{2}$ given in eq. (\ref{j1j2})) yields
\bea
Q & = & 2 \phi_H \, (M + 2U_H N) \, - \, 2 U_H P \, ,
\nonumber\\
P & = & 2 \psi_H \, (M + 2U_H N) \, + \, 2 U_H Q \, .
\label{rel-5}
\eea
For vanishing NUT charge eq. (\ref{rel-6}) gives $U_H = 0$.
Otherwise we can solve for $U_H$ and use the result in 
eqs. (\ref{rel-5})
to obtain the explicit expressions for the potentials 
$\phi_H$ and $\psi_H$ in trems of the charges $M$, $N$, $Q$ and $P$.
One finds
\bdm
\phi_H \, = \, \frac{1}{2 N} \left( 
\frac{N Q - M P}{\sqrt{M^2 + N^2}} + P \right) \, , \; \; \; \; \; 
\psi_H \, = \, \frac{1}{2 N} \left(
\frac{N P + M Q}{\sqrt{M^2 + N^2}} - Q \right) \, ,
\edm
\be
U_H \, = \, \frac{1}{2 N} \left(
\sqrt{M^2 + N^2} \, - \, M \right) \, .
\label{res-1bis}
\ee
We may eventually use these formulae to eliminate
the potentials from the 
relations (\ref{rel-3}) and (\ref{rel-4}). A short
calculation gives 
$2D(M^2 + N^2) = 2NPQ + M(Q^2 -P^2)$ and
$2A(M^2 + N^2) = 2MPQ - N(Q^2 -P^2)$. Hence, 
\be
2 \, M_c \, D_c \, = \, (Q_c)^2 \, , 
\label{res-2}
\ee
where the complex charges $M_c$, $Q_c$ and $D_c$ are defined by
\be
M_c \, \equiv \, M + i N \, , \; \; \; \;  
Q_c \, \equiv \, Q + i P \, , \; \; \; \;  
D_c \, \equiv \, D + i A \, .
\label{defs-complex}
\ee
We have now exhausted all information from
the additional conservation laws following from the
coset structure. The only equation which was not
used yet is the Smarr formula (\ref{rel-2}). 
Substituting the horizon-values
(\ref{res-1bis}) for the potentials into the Smarr formula
(\ref{rel-2}), we obtain the following 
expression for $M_H$ in terms of the charges:
\be
M_H \, = \, \frac{2 \, \mid M_c \mid ^{\, 2} 
\, - \, \mid Q_c \mid ^{\, 2}}{ 2 \, \mid M_c \mid} \, . 
\label{res-2bis}
\ee
Taking the square of this formula (and using eq. (\ref{res-2})
to eliminate the $\mid Q_c \mid^4$-term) 
finally yields the desired expression,
$M_H^2 = \mid D_c \mid^2 + \mid M_c \mid^2 - \mid Q_c \mid^2$,
that is,
\be
\left(\frac{1}{4 \pi} \kappa {\cal A}\right)^2 \, + \, 
Q^2 \, + \, P^2 \, = \, M^2 \, + \, N^2 \, + \, D^2 \, + \, 
A^2 \, .
\label{res-end} 
\ee

For $M_H = 0$, the above formulae have been obtained for
various spherically symmetric BPS solutions of the 
EM and EMDA equations (see, e.g., \cite{CG-96}, \cite{GL-96} 
and references therein). 
For spherically symmetric configurations with 
nondegenerate horizons ($\kappa \neq 0$),
eq. (\ref{res-end}) was obtained by Breitenlohner
{\it et al.} \cite{BMG-88}, 
where the term on the l.h.s. was not specified. More 
recently, Gibbons {\it et al.} \cite{GKK-96} were able to 
establish the {\em full\/} relation (\ref{res-end}) for
spherically symmetric solutions, using the generalized first 
law of black hole thermodynamics. A version of eq. 
(\ref{res-end}) which also includes the rotation parameter 
was derived by Gal'tsov and Kechkin \cite{GK-94} for the 
dilaton-axion-Kerr-NUT dyon solution. The latter was 
constructed from the Kerr-NUT metric, using the symmetries of
the target space. The relation was also derived by applying
generalized Harrison and Ehlers transformations to the seed 
Schwarzschild solution \cite{GL-96}, \cite{GL-EH}.

The above derivation shows that the generalization 
(\ref{res-end}) of the Bogomol'nyi equation holds for arbitrary 
stationary, asymptotically flat (or asymptotically NUT) 
solutions of the EM and EMDA equations. 
The non-negative term which transforms the 
inequality $M^2 + N^2 + D^2 + A^2 - Q^2 - P^2 \geq 0$ into
an equality is found to be 
$(\frac{1}{4 \pi} \kappa {\cal A})^{2}$.
Although we have established the above results by using 
explicit representations of the EM and EMDA cosets, we 
expect them to hold in the general case as well. More precisely,
we conjecture that the Hawking temperature of
all stationary, asymptotically flat 
(or asymptotically NUT) black holes with massless 
scalars and Abelian vector fields is given by
\be
T_H \, = \, \frac{2}{{\cal A}} \, \sqrt{\sum (Q_S)^2 \, - \, 
\sum (Q_V)^2} \, ,
\label{HTEMP}
\ee
provided that the field equations assume the form
(\ref{i-1}), (\ref{i-2}) and
$\Phi$ is a map into a {\em symmetric space}, $G/H$.
Here $Q_S$ and $Q_V$ denote the charges of the scalers
(including the gravitational ones)
and the vector fields, respectively.

\section{Acknowledgments}

It is a pleasure to thank Gary Gibbons for a stimulating
discussion, during which he conjectured that the generalized
Smarr formulae (which I previously
obtained for the EM system by different means \cite{MH-B})
should be related to the sigma-model structure 
of the stationary field equations. I am also grateful
to O. Brodbeck, D. Gal'tsov, N. Straumann and M. Volkov
for helpful suggestions and discussions. 
This work was supported by the Swiss National Science Foundation.

\appendix

\section{}\label{appendix}

In this Appendix we recall some identities for Killing fields.
Throughout, $k$ will denote a timelike Killing field ($1$-form)
with norm $V$ and twist ($1$-form) $\omega$:
\be
V \, \equiv \, - \sprod{k}{k} \, , \; \; \; \; \;
\omega \, \equiv \, \frac{1}{2} \ast \left( k \we dk \right) \, .
\label{A-1}
\ee
The Lie derivative of an arbitrary $p$-form $\alpha$ with respect 
to a Killing field commutes with the Hodge-dual, i.e.,
\be
L_{k} \ast \alpha \, = \, \ast \, L_{k} \alpha \, , \; \; \; 
\mbox{where} \; \; L_{k} \, = \, i_{k} \circ d + d \circ i_{k} \, .
\label{A-2}
\ee
The operator $i_{k}$ denotes the interior product (derivative)
assigning to $\alpha$ the $(p-1)$-form 
$(i_{k} \alpha)_{\mu_{2} \ldots \mu_{p}} \equiv 
\alpha_{\mu \mu_{1} \ldots \mu_{p}} k^{\mu}$. The latter is also
obtained from the exterior product of the dual of $\alpha$
with $k$:
\be
i_{k} \alpha \, = \, -\ast (k \we \ast \alpha) \, , \; \; \; \; \; 
i_{k} \ast \alpha \, = \, \ast (\alpha \we k) \, ,
\label{A-3}
\ee
where the second identity is obtained from the first one by
replacing $\alpha$ with its dual and using
$\ast^{2} \alpha = -(-1)^{p} \alpha$. For an invariant $1$-form
$\alpha$, $L_{k} \alpha = 0$, eqs (\ref{A-2}) and
(\ref{A-3}) imply
$d \ast (k \we \alpha) = - d (i_{k} \ast \alpha) =
i_{k} d \ast \alpha$, and hence
\be
d \ast (k \we \alpha) \, = \, -(\cod \alpha) \ast k \, , \; \; \; 
\mbox{if} \; \; L_{k} \alpha \, = \, 0 \, ,
\label{A-4}
\ee
where $\cod \equiv \ast d \ast$ is the co-derivative operator.
The above formula also provides one with a coordinate invariant 
formulation of Stokes' theorem for stationary (but not
necessarily static) spacetimes.

The Frobenius theorem -- implying that the hypersurfaces of 
constant $V$ are orthogonal to $k$ if and only if the twist
$\omega$ vanishes -- is recovered from the identity
\be
d \left( \frac{k}{V} \right) \, = \, 
2 \ast (k \we \frac{\omega}{V^{2}}) \, = \, 
- \frac{2}{V^{2}} i_{k} \ast \omega \, .
\label{A-5}
\ee
(This is obtained from eqs. (\ref{A-1}) and (\ref{A-2}), 
which yield
$2 i_{k} \ast \omega = i_{k} (k \we dk) = -V \, dk - k \we dV$,
since $i_{k} dk = -di_{k}k = dV$.) Applying
the exterior derivative to the above identity and 
using eq. (\ref{A-2})
yields $0 = i_k d \ast (\omega / V^2)$ and thus 
(since $d \ast (\omega / V^2)$ is a $4$-form) 
\be
\cod \left( \frac{\omega}{V^2} \right) \, = \, 0 \, .
\label{A-6}
\ee
The identities (\ref{A-5}) and (\ref{A-6}) also imply the 
following: Let $\Omega$ be a closed $2$-form which is invariant
with respect to the isometry group generated by the Killing
field $k$. Then there exist (locally) a function $f$
and a (current) $1$-form $j$, such that
\be
\cod j \, = \, 0 \, , \; \; \; \mbox{where} \; \; 
j \, = \, \frac{i_k \ast \Omega}{V} 
\, - \, 2 \, f \, \frac{\omega}{V^2} 
\; \; \; \mbox{and} \; \; 
df \, = \, i_k \Omega \, .
\label{A-7}
\ee
First, by virtue of eq. (\ref{A-2}), $L_k \Omega = 0$ and 
$d \Omega = 0$ imply $d (i_k \Omega) = 0$ and thus 
the local existence
of a potential $f$, such that $df = i_k \Omega$. Using this and the
identity (\ref{A-5}) gives
$d (\frac{k}{V} \we \Omega)$ $=$ 
$-2 \Omega \we i_k \ast \frac{\omega}{V^2}$ $=$
$2 df \we \ast \frac{\omega}{V^2}$ $=$
$2 d (f \ast \frac{\omega}{V^2})$, where we have taken
advantage of the identity (\ref{A-6}) in the last step.
Using eq. (\ref{A-3}) in the first term proves the above formula.

As an application of eq. (\ref{A-7}) one can write the Maxwell
equations in the presence of a Killing field in the form of 
conservation laws for two current $1$-forms. The Bianchi
identity and the Maxwell equation imply that both the 
electromagnetic $2$-form $F$ and its dual $\ast F$ are closed.
Moreover, assuming that $F$ is stationary, $L_k F = 0$,
eq. (\ref{A-2}) implies that $\ast F$ is stationary as well,
$L_k \ast F = 0$. Hence, we can either choose $\Omega = F$
or $\Omega = \ast F$ in eq. (\ref{A-7}). Introducing the potentials
$\phi$ and $\psi$, defined by $-d \phi = i_k F$ and
$d \psi = i_k \ast F$, respectively, the stationary Maxwell 
equations become
\bea
\cod j_{P} & = & 0 \, , \; \; \; \mbox{where} \; \; 
j_{P} \, = \, 
\frac{d \psi}{V} + 2 \phi \frac{\omega}{V^2} \, ,
\nonumber\\
\cod j_{Q} & = & 0 \, , \; \; \; \mbox{where} \; \; 
j_{Q} \, = \,
\frac{d \phi}{V} - 2 \psi \frac{\omega}{V^2} \, .
\label{A-8}
\eea


In the presence of a (timelike) Killing field, the Ricci tensor
can be reduced with respect to the projection metric 
$\pmetric = V \gtens + k \otimes k$. The 
$\Rtens(k, \, \cdot \, )$-components can also be obtained from the
Ricci-identity,
\be
-\Delta k \, = \, \cod d k \, = \, 2 \, \Rtens(k) \, ,
\; \; \; \mbox{with} \; \; 
R(k)_{\mu} \, \equiv \, R_{\mu \nu} k^{\nu} \, , 
\label{A-9}
\ee
by expressing the Laplacian of $k$ in trems of $V$ and $\omega$.
(For a Killing $1$-form one has $\cod k = 0$ 
and therefore $-\Delta k = (\cod d + d \cod) k = \cod dk$.)
To this end, one uses eq. (\ref{A-5}) in the form
\be
\ast dk \, = \, i_k \ast \frac{dV}{V} \, - \, 2 \,
\frac{k}{V} \we \omega \, .
\label{A-10}
\ee
Applying the exterior derivative to this identity
and taking advantage of
eq. (\ref{A-2}) in the first and of eq. (\ref{A-5}) 
in the second term gives
\bdm
d \ast dk \, = \, -i_k d \ast \frac{dV}{V} + 
\frac{4}{V^2} i_k \ast \omega \we \omega +
2\, \frac{k}{V} \we d \omega \, .
\edm
Now using $i_k \omega = 0$, $i_k \ast 1 = \ast k$ and the
Ricci identity (\ref{A-9}) yields the desired result,
\be
2 \, \Rtens(k) \, = \, -\Delta k \, = \,
\left[ \cod \left( \frac{dV}{V} \right) - 4 \, 
\frac{\sprod{\omega}{\omega}}{V^2} \right] \, k \, + \,
2 \ast \left( \frac{k}{V} \we \omega \right) \, .
\label{A-11}
\ee
Since the last $1$-form is orthogonal to $k$, we immediately
find
\bea
2 \, \Rtens(k,k) & = &
-V \, \cod \left( \frac{dV}{V} \right) + 4 \,
\frac{\sprod{\omega}{\omega}}{V} \, ,
\label{A-12} \\
k \we \Rtens(k) & = &
- \ast d \omega \, , 
\label{A-13}
\eea
where the second equation implies
$\Rtens(k,X) = \frac{1}{V} (\ast d \omega)(k,X)$, for any
vector field $X$ orthogonal to $k$.
Finally, we also recall the projection formula
for the remaining components of the Ricci tensor
\cite{BG-71} (see, e.g., \cite{KSMH-80}). For
$X$ and $Y$ orthogonal to $k$ one finds
\be
\Rtens(X,Y) \, = \, \frac{1}{V} \Rtens(k,k) \, \gtens(X,Y) \, + \, 
\RPtens(X,Y) \, - \, \frac{1}{2 V^2} \left(
dV \otimes dV + 4 \, \omega \otimes \omega \right)
\label{A-14}
\ee
where $\RPtens$ denotes the Ricci tensor obtained from the
metric
$\pmetric = V \gtens + k \otimes k$.

\end{document}